\documentclass{IEEEtran}
\usepackage{color,soul}
\usepackage{graphicx,url}
\usepackage{multirow}
\usepackage{amsmath,amssymb}
\usepackage{graphicx}
\usepackage{url}

\usepackage{multicol,multirow}
\usepackage{booktabs}
\usepackage{subfigure}
\usepackage{flushend}
\usepackage{float}
\usepackage{nth}
\usepackage{todonotes}

\usepackage{array}
\usepackage{flushend}

\begin{document}
\title{Android Malware Characterization using Metadata and Machine Learning Techniques}

\author{\authorblockN{Ignacio Mart\'{i}n\authorrefmark{1}, Jos\'{e} Alberto Hern\'{a}ndez\authorrefmark{1}, Alfonso Mu\~noz\authorrefmark{2}, Antonio Guzm\'{a}n\authorrefmark{2}\\}
\authorrefmark{1}Universidad Carlos III de Madrid, Spain\\ Email: {\{ignmarti, jahgutie\}}@it.uc3m.es\\
\authorblockA{\authorrefmark{2}Telef\'{o}nica Digital Identity \& Privacy, Spain \\ Email:\{Alfonso.Munoz, Antonio.Guzman\}@11paths.com}
}
\maketitle

\begin{abstract}
Android Malware has emerged as a consequence of the increasing popularity of smartphones and tablets. While most previous work focuses on inherent characteristics of Android apps to detect malware, this study analyses indirect features and meta-data to identify patterns in malware applications. Our experiments show that: (1) the permissions used by an application offer only moderate performance results; (2) other features publicly available at Android Markets are more relevant in detecting malware, such as the application developer and certificate issuer, and (3) compact and efficient classifiers can be constructed for the early detection of malware applications prior to code inspection or sandboxing.
\end{abstract}


\maketitle
\section{Introduction and Motivation}

The mobile market industry has explosively grown in the last decade. According to latest estimates, the number of smartphone users has reached 2 billion at the beginning of 2014, and is expected to grow up to more than 2.50 billion in 2018 \footnote{See: \url{http://www.emarketer.com/Article/2-Billion-Consumers-Worldwide-Smartphones-by-2016/1011694}, last access Nov 2016}.

Android has positioned itself as the leading operating system in the Smartphone industry, accounting for more than 86.8\% of devices by the end of 2016\footnote{See \url{http://www.idc.com/prodserv/smartphone-os-market-share.jsp}, last access Mar 2017}. Indeed, one key for its success is that the Android platform is open to any developer, individual or enterprise, who is able to easily design new applications and services and upload them to any of the Android markets available, namely Google Play Store, Amazon Appstore, Samsung Galaxy Apps, etc. At the time of writing, it is estimated that nearly 2.7M applications are uploaded at Google Play, while new applications are uploaded at a pace of more than 60k per month\footnote{See \url{http://www.appbrain.com/stats/number-of-android-apps}, last access Mar 2017}.

Unfortunately, the popularity of Android and its ease for develop and upload any app has side effects. In this light, Android has become one of the most valuable targets for malware developers. An extensive taxonomy of Android malware applications, where up to 49 malware families have been identified, can be found in~\cite{zhou12:_dissec_android_malwar}.

The ability to early detect malicious Android applications is vital to enhance user security, since Android apps can be tagged, reported and removed from the market and their signatures blacklisted. This is a classification problem and, therefore, many authors have attempted the application of machine learning to different feature sets. 

Consequently, machine learning has been profoundly studied, and a survey of techniques may be found in \cite{MLstudy}. For instance, the authors in \cite{drebin} gather features from application code and manifest (permissions, API calls, etc) and use Suport Vector Machines (SVMs) to identify different types of malware families. The authors in \cite{bayesianAnalysis} analyse bayesian-base machine learning techniques for Android malware detection. In~\cite{6298824}, the authors  use permissions and control flow graphs along with Suport Vector Machines (SVMs) to differentiate malware from good applications ("goodware" in what follows).  The authors in~\cite{6735264} uses API calls and permissions as features to train SVMs and Decision Trees. Androdialysis~\cite{AndroDialysis} explores the intents of each application as features for the classification task. Yerima et al~\cite{APICallsEnsembles} try different algorithms over API calls and command sets and shows promising results for ensemble methods, such as random forest.

In general, Android permissions have been extensively studied under the assumption that these are critical in identifying most malware,  see~\cite{puma,aung2013permission,empiricalAnalysis,analysisPermCongress}. Actually, in \cite{puma} the authors show that malware uses less permissions than goodware. 

The authors in~\cite{6675404} attempt to detect malware by inspecting other application run-time parameters, such as CPU usage, network transmission and process and memory information. Mas'ud et al~\cite{6847364} also include Android system calls into analysis. Furthermore Elish et al~\cite{profilingUsers} propose a single-feature classification system based on user behaviour profiling. Droidchain authors~\cite{droidChain} propose a novel model which analyses static and dynamic features of applications assuming different malware models.

  In a different approach, the authors of {\cite{massVetting}} design a differential-intersection analysis technique to identify repackaged versions of popular applications, which is a common way to disguise malicious applications, showing good performance.

Concerning malware detection systems, there exists two main trends: (1) online services which aim to provide efficient and lightweight solutions to cope with malware detection from the mobile device and (2) offline services to engage in fast analysis of enormous amounts of applications in order to mark potentially harmful code, either for removal or extended inspection.  In this light, several authors have explored both trends, obtaining results such as the systems exposed in~\cite{drebin,andromaly,secloud} which provide online solutions to inform or warn the user on the device or more general, hardware-dependent systems such as~\cite{riskRanker,droidapiminer} which are huge scalable systems capable of dealing with huge amounts of applications at once, enabling fast and cheap detection mechanisms for entities like application markets to improve the quality of their apps. \cite{malwaretrends} surveys extensively the types and works regarding malware detection system.

In addition, obtaining as much information as possible on threats and other undesired applications is really necessary, and various authors propose methodologies and systems to collect diverse and huge amounts of data. For example, Burguera et al~\cite{crowdroid} propose a framework for collecting application trace and identify uncommon behaviours of common applications. Moreover, the authors of~\cite{signatureBasedMobile,droidAnalytics} propose a system to gather signatures and malware information automatically.

In fact, a good deal of information is already available at Google Play and can be used to identify patterns not yet pointed out in previous work. Elements like developer name, categories or votes have not been used to the best of our knowledge in malware detection yet. Such meta-data provides a good starting point to develop a lightweight malware detector which does not require performing behaviour analysis and provides a fast first-stage notion on whether an application "behaves suspiciously" (shows malware patterns) or not. However, very few studies have analysed any subset of this information: only the authors of~\cite{Nannen:Thesis:2003}  performed sentiment analysis on the comments made by users regarding Android applications. 

%a good starting point to develop a lightweight malware detector which does not require performing behaviour analysis and provides a fast first-stage notion on whether an application "behaves suspiciously" (shows malware patterns) or not.

To this end, this work focuses on the analysis of such indirect features and their ability to unveil malware. We analyse meta-data to find only a subset of features which have proven predictive power and use them to develop and test different machine learning models. 

%Such features combined with powerful machine learning algorithms provide a good framework to help Android users identify suspicious malware without exhaustive application analyses like sandboxing or code inspection.

The remainder of this work is organised as follows: Section~\ref{sec:dataset} describes the dataset under study, including number of applications and types of features analysed. Section~\ref{sec:methodology} is explains to the methodology, whereas section~\ref{sec:results} reports the experiments and results obtained. Finally, section~\ref{sec:conclusions} concludes this work with a summary of the findings.

\section{Dataset description and pre-processing}
\label{sec:dataset}

The dataset used in this study comprises around 140K Android applications collected from Google Play Store during 2015. This dataset has been obtained using the Tacyt cyber-intelligence tool developed internally at Eleven Paths (Telef\'{o}nica Group, see Acknowledgements for further details). For each application, we have extracted not only intrinsic features of the Application PacKage File (apk), e.g. size in bytes or list of permissions used, but also other meta-data available at Google Play, including that related to the application developer or the number of votes or average star rating. Some of these features are numeric (e.g. application size, average rating), while others are categorical (e.g. whether an application belongs to a certain category or not). Next section overviews the features derived from such data, some of them will reveal extremely powerful in identifying potential malware.

%not only intrinsic features of the Application PacKage File (apk) but also other meta-information available at Google Play. Some of these features are numeric (e.g. application size, average rating), while others are binary (e.g. whether an application belongs to a certain category or not). 

\subsection{Intrinsic application features} 
These relate to concise application information, including its size (bytes), application category, number of images and files used by the application, etc. This group comprises 15 features. 

Other intrinsic features considered in the analysis include the permissions used by each apk. There are over 29K different permissions used by the applications in our dataset; most popular ones are: 
\begin{itemize}
\item android.permission.internet (found in 96.07\% of apps)
\item android.permission.access\_network\_state (91.15\%)
\item android.permission.read\_external\_storage (54.5\%)
\item android.permission.write\_external\_storage (54.12\%)
\item android.permission.read\_phone\_state (39.81\%)
\end{itemize}

Many permissions appear only once in the dataset as they are often self-defined permissions. Thus, the binarized permission features comprise a very-sparse high-dimensional matrix. In these cases, feature hashing~\cite{featureHashing} is an effective strategy for dimensionality reduction; it works by grouping applications according to some hash functions. We will leverage the hashing trick in the paper to increase the number of Intrinsic application features in a smaller number that just adding permissions as they are.

%applying a hash function to the features and using their hash values as indices directly. The number of hash functions must be large enough to capture the patterns in the permissions set whilst minimising the number of collisions. We shall use 512 hash functions in our experiment section. Thus, intrinsic features comprise a total of 15+512 variables in total.

\subsection{Social-related features} 
These are 7 features and involve feedback collected from users in the market. As Google Play is strongly connected with the social network Google+, features like total number of votes or average rating are provided. For each possible ratings (1, 2, 3, 4 and 5 stars) we acquire the number of votes given. Then, it is possible to easily compute the mean average of any application in the market as well as the total number of votes for that application.

\subsection{Entity-related features: Developers and Certificate Issuers}
Android markets often provide information about the application developers (name, email address, website, etc), and the certificate information of the application signature (expedition or expiration dates, issuer or subject names, etc). 

Within data, there are around 45K different developer names and 40K certificate issuer names. Following~\cite{reputation}, we have created two new features called developerRep and issuerRep which account for the percentage of applications that each developer and certificate issuer have tagged as malware. The reader must note that Google Play allows self-signed applications, i.e. applications where the issuer is the same as the developer. 

As a result, in many cases, the issuer of a certificate and the developer may be the same entity. However, their reputations may change, as many issuers may not only sign their own applications and not all developers self-sign their applications (and even if they do, they use different accounts)

%These two features have revealed extremely powerful as noted in the experiments section.

\subsection{Malware detection attributes} 

Once downloaded, all applications have been inspected for malware using the VirusTotal web service (free Online Virus, Malware and URL Scanner, available at:  \url{http://www.virustotal.com/}, last access Feb 2017). VirusTotal checks each application against a large number of malware engines,  producing a binary result (malware/goodware) per engine (McAfee, AVG, VIPRE, TrendMicro, etc.). In our dataset, around 50\% of the applications have been declared as malware by at least one of these engines. 

Concerning the number of detectors per malware application, a zipf-like behaviour is observed, i.e. most malware applications are only detected by a single antivirus (AV) engine, while a few number of malware applications are detected by many AV engines. In particular, 25\% of the malware applications are detected by 1 AV engines or less (1st Quartile),  50\% are detected by 2 AV engines or less (median) and 75\% malware applications are detected by 4 AV engines or less (3rd Quartile). We shall use the label "isMalware" (TRUE/FALSE) to denote whether an application is tagged as Malware or not.

\begin{figure}[!htbp]
\includegraphics[width=\columnwidth]{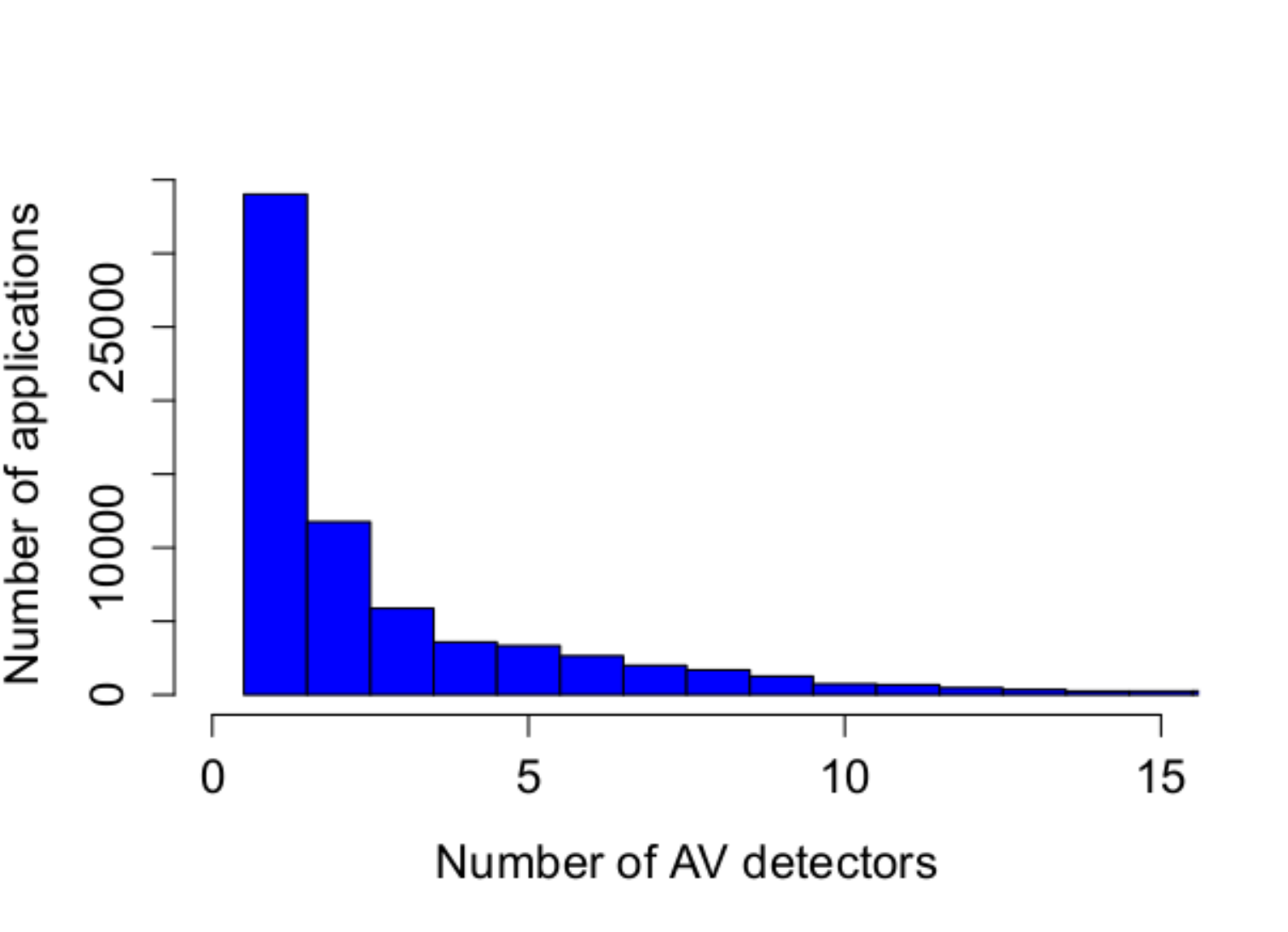}
\caption{Histogram of AV detectors per malware application.}
\label{fig:histogram}
\end{figure}

Fig.~\ref{fig:histogram} shows in a histogram the frequency of each application detection count. The zipf-like behaviour is clear in the picture, as most applications are only detected by a single engine (34,025 applications), while the average detection count is 3. Furthermore, there is one application detected as malware by 53 AV detectors.

Due to this disparity and disagreement among AVs, we will consider the aforementioned quantiles (1,2 and 4 detections) as different thresholds to establish ground truth rules within the detection scheme.

%25\% of the malware applications are detected by 1 AV engines or less (1st Quartile),  50\% are detected by 2 AV engines or less (median) and 75\% malware applications are detected by 4 AV engines or less (3rd Quartile). We shall use the label "isMalware" (TRUE/FALSE) to denote whether an application is tagged as Malware or not.
\section{Methodology and Data Analysis}
\label{sec:methodology}

\subsection{Initial approach}

\begin{figure*}[!htb]
\centering
\subfigure[Number of Downloads]{
\includegraphics[width=0.3\textwidth]{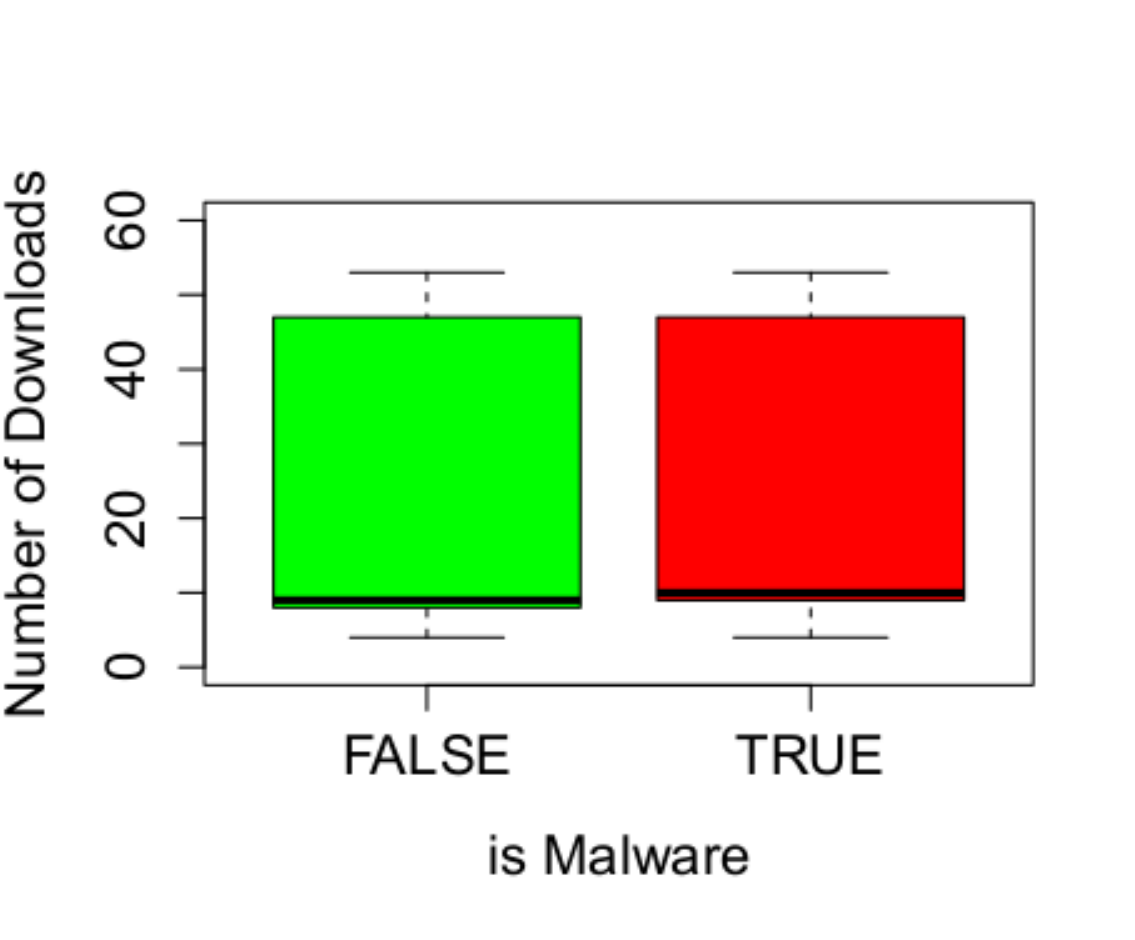}
}
\subfigure[Number of Days in Google Play]{
\includegraphics[width=0.3\textwidth]{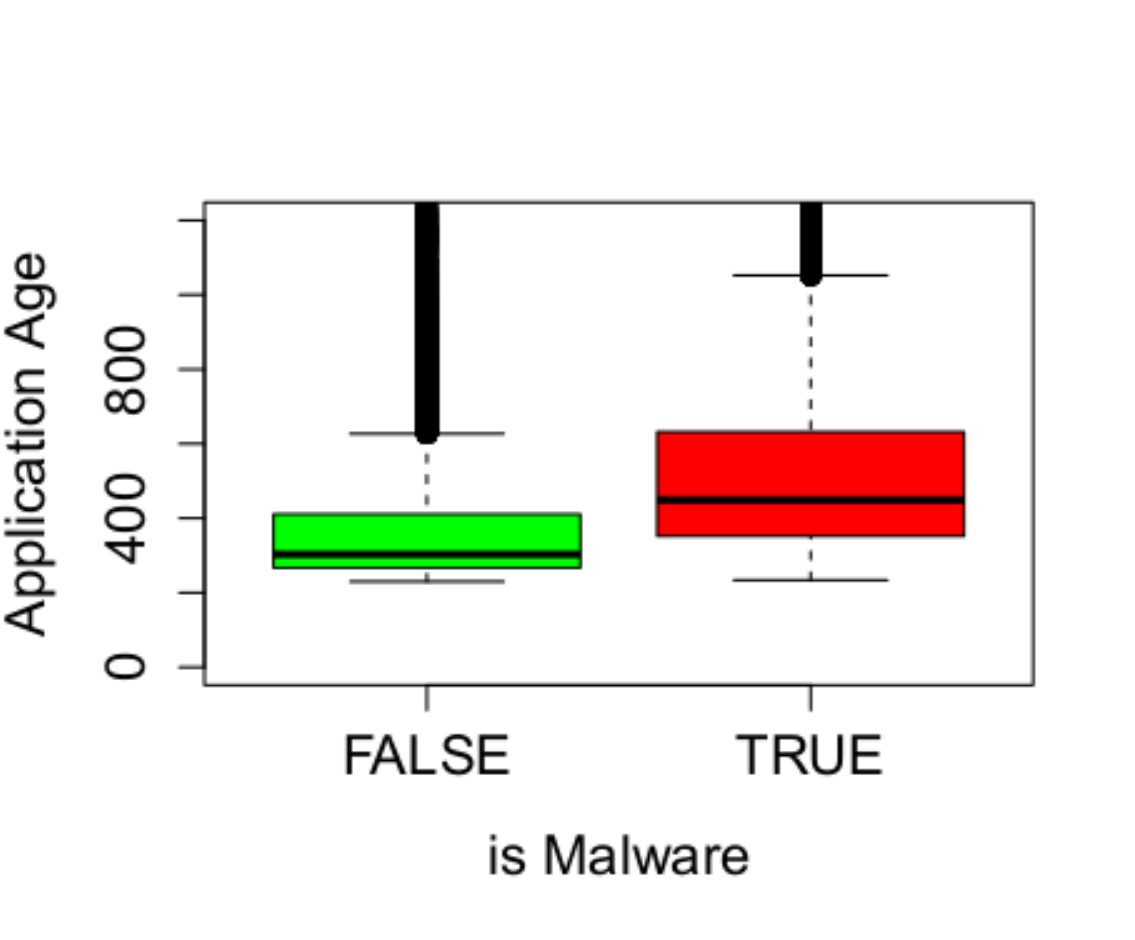}
}
\subfigure[Associated Developer Reputation]{
\includegraphics[width=0.3\textwidth]{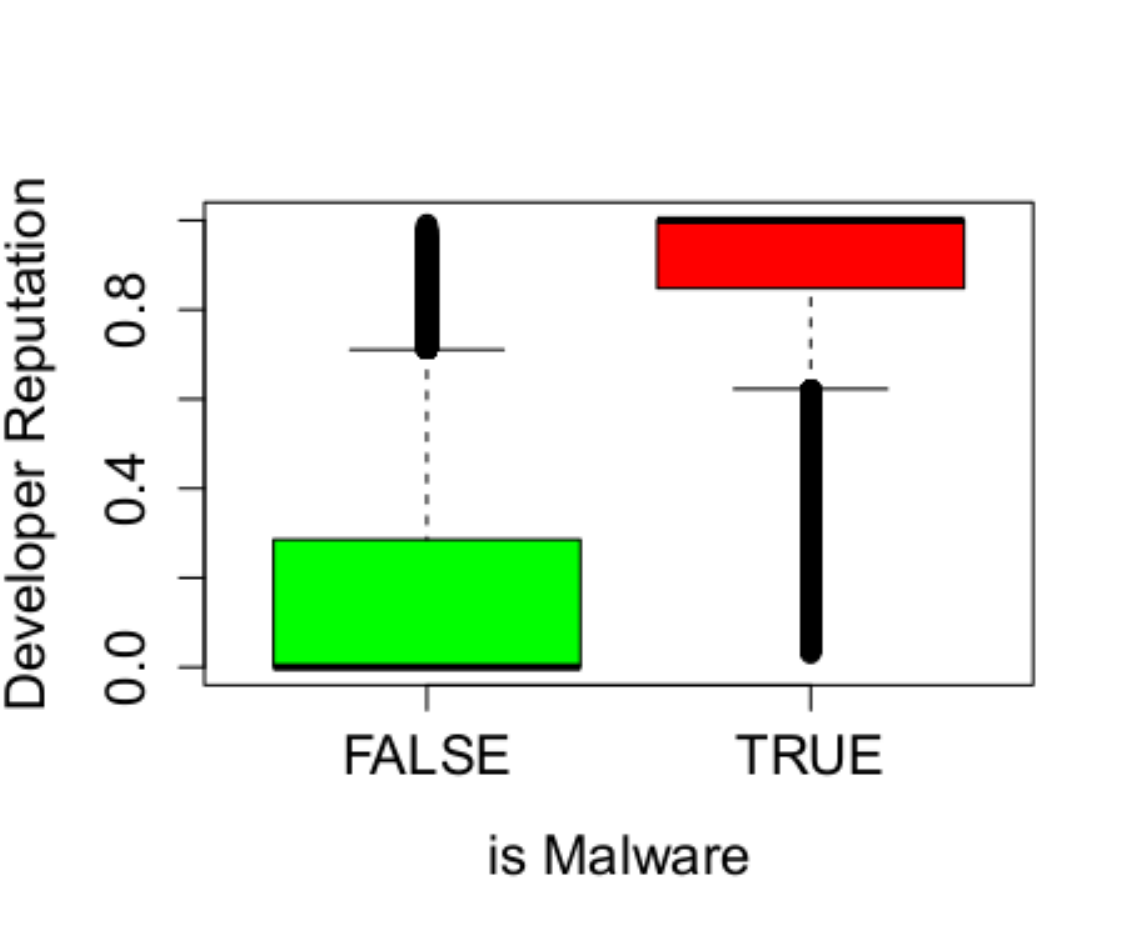}
}

\caption{Goodware/Malware boxplot comparison for three features:Number of downloads, number of days since the application was uploaded and developer reputation}
\label{fig:featuresboxplot}

\end{figure*}

Feature selection is key to reduce complexity and improve performance. We expect some features to have more predictive power than others, as noted in Fig.~\ref{fig:featuresboxplot}. In this figure, three boxplots for malware/goodware classes are shown for three sample features: the number of times the application has been downloaded from the market (left), the time the application has been in Google Play (centre) and the developer reputation (right). 

As observed, downloads is not a very useful feature, since both goodware and malware show  similar 25-percentile (around 10) as well as 75-percentile (48), values. Concerning the number of days in Google Play (centre), the 25-, 50- and 75- quantile measures of malware differ from goodware, showing some predictive power. Finally, developers reputation (right) clearly reveals that malware developers tend to develop more malware while goodware developers create almost no malware. 

%This intuition is later on confirmed with a number of feature selection algorithms applied to the dataset.

\subsection{Classification models and performance evaluation}

In a binary classification problem, we are often given a training set with labeled data $\{X_i,y_i\}_{i=1}^{N_{tr}}$, where $y_i\in\{0,1\}$ and $X_i$ is a vector containing the values of $P$ predictors or features, namely, $X_i=(x_{i1},\ldots,x_{iP})$. In our case, the labels $y$ refer to the categoric variable "isMalware", whereas the predictors $X_i$ comprise 512 feature hashes of permissions, 15 intrinsic features, 7 social-related features and reputations.

%"developerRep" and "issuerRep".

Machine-Learning algorithms are in charge of constructing a function $g(X)$ from the training set that separates the two classes with minimum error. In our experiments, we have used \emph{Logistic Regression (LR)}, \emph{Support Vector Machines (SVMs)} and \emph{Random Forests (RF)} as three well-known  supervised learning algorithms 
%(see~\cite{statlearning} as a good overview of statistical learning). 

Once a model is obtained, the following stage consists on testing its ability to predict the result of unobserved data samples, i.e. evaluate the model's generalisation capabilities. Ten-fold cross-validation has been used to adjust resulting models and evaluate test error, with well-known metrics: \emph{Receiver Operating Characteristic (ROC) curves} and the \emph{Area Under ROC Curve (AUC-ROC)}, \emph{Precision}, \emph{Recall} and \emph{F1-score}. 

It is worth recalling that the ROC curve compares the False Positive Rate (FPR) vs True Positive Rate (TPR), and the AUC measures the integral of the ROC curve, being unity the highest possible value. In addition, Precision measures how many of the applications tagged as malware are indeed malware, while Recall measures how many true malware applications the model detects from the total. In other words:
$$\textrm{Prec} = \frac{TP}{TP+FP} \quad\quad \textrm{Recall} = \frac{TP}{TP+FN}$$
where (TP, FP, TN, FN) refer to True/False Positive/Negative respectively. Finally, the F1-score trades off precision and recall by computing their geometrical mean.

\subsubsection{Validation and Significance}

Ten-fold cross-validation consists on splitting the entire dataset in 10 chunks of equal size and perform 10 iterations over them, selecting at each turn a different chunk to be the testing set and the reminding ones to be the training. Using this method, one can perform hyper-parameter tuning, but also provide results with statistical significance (i.e. robust results which do not depend on the selection of training/test instances).

\subsection{Feature selection}

Some features are critical in the discrimination between good/malware while others are not, either due to correlation or small predictive power. For selecting from those features, we have used the following methods:

%because they show a similar pattern for both good/malware or are highly correlated with other features thus adding little value. Different techniques have been used to extract only the most relevant features, including:

\subsubsection{Pearson's Chi Squared test} Statistical test used to determine whether any difference among variables occurs by chance or there is indeed a statistical relation.

\subsubsection{Entropy-based methods} In information theory, entropy measures the amount of unknown information a certain source provides. The following measurements are considered:
\begin{itemize}
\item \textit{Information Gain (IG)} or mutual information between a feature $X_i$ and the outcome $y$.
\item \textit{Gain Ratio} is the result of dividing the information gain by the intrinsic information of the feature, aiming to reduce bias towards features with high information gain value on its own rather than a good relationship with the output variable $y$. 
\end{itemize}

\subsubsection{Random Forest importance} or contribution of its nodes, in particular the Mean Decrease in Node Impurity (MDNI), which measures the inequality among nodes within a Random Forest.

For further reference of machine learning and statistical methods for data analysis, please refer to \cite{statlearning}.

\section{Experiments and results}
\label{sec:results}

In the experiments, we have used the well-known R open-source statistical software, along with a number of libraries for machine learning and feature selection (MASS, randomForest, kernlab and glmnet). From the original dataset, we have built nine different subsets with different compositions. Concisely, for each subset we vary either the amount of malware it contains (2\%, 25\% or 50\% of malware over the total) and the threshold used for considering an application as malware (1, 2 or 4 AV detectors). As an example, we shall refer to the (1-AV, 25\%) malware dataset as a dataset that contains 25\% malware and 75\% goodware applications where malware is randomly selected among all applications whereby at least 1-AV detector was fired.

Each of these subsets include an amount of 50K applications, except the 50\%-4AV dataset which only contains 36K samples due to the lack of malware applications meeting the conditions to be considered malware.

\subsection{Predictive power of permissions}

As noted in the introduction, several researchers have studied the permissions used by an application and their ability to detect malware. For instance, the authors in~\cite{droidPermissionMiner} achieve F1-score values in the range of 0.6 to 0.8.

In order to evaluate the effects that feature hashing has on permissions, we try different hashing spaces (32, 64, 128, 256, 512, 1024 and 2048 hashes)to evaluate the feature amount- performance trade off. To measure performance, we run 10-fold cross-validation for threshold tuning in a logistic regression algorithm and compute different AUC (Area Under the Curve) measurements for each of the hashing spaces.

\begin{figure}[!htbp]
\includegraphics[width=\columnwidth]{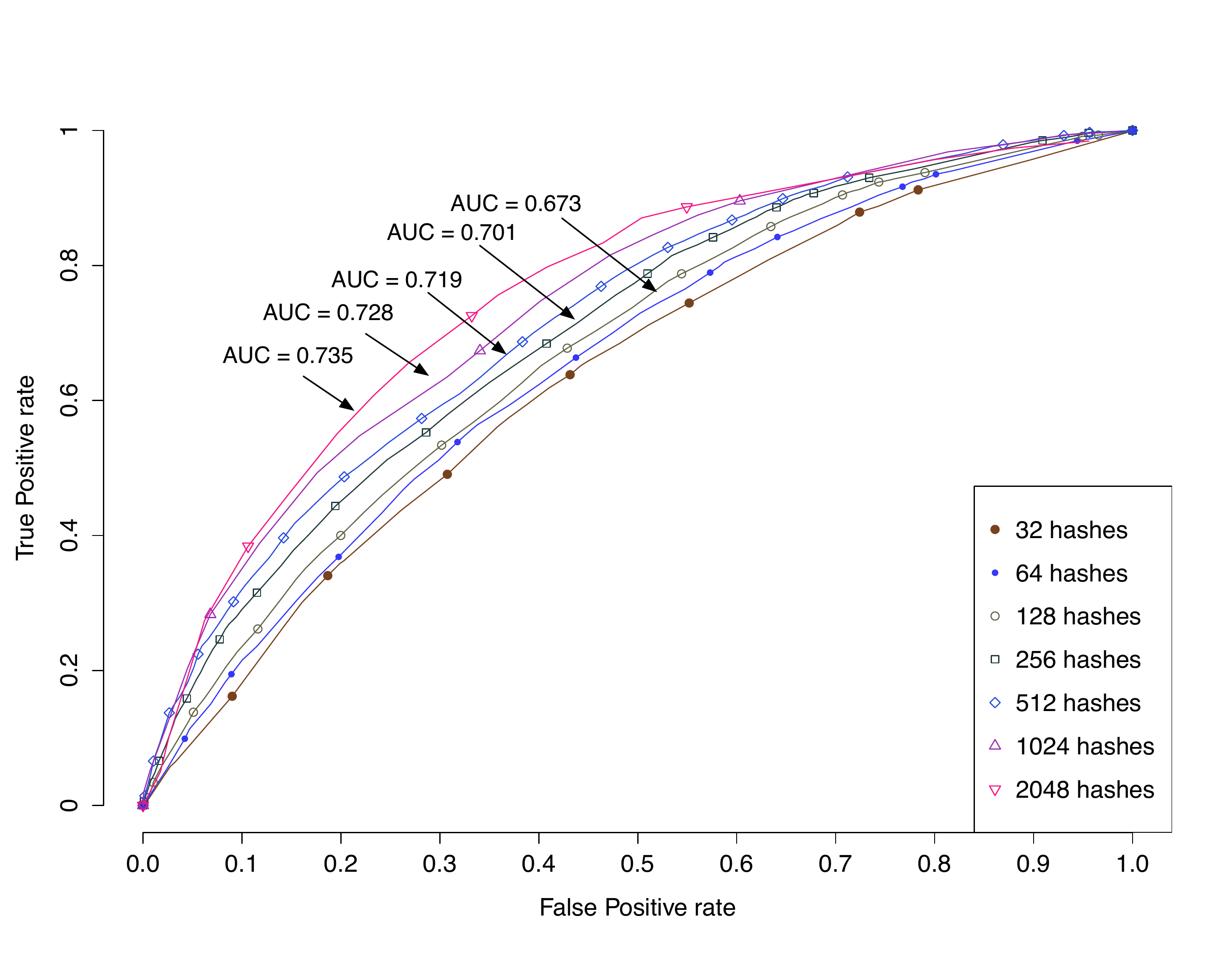}
\caption{ROC curve for malware detection using feature hashing on permissions only.}
\label{fig:featureHashes}
\end{figure}

In our case, Fig.~\ref{fig:featureHashes} shows the ROC curve and AUC-ROC values using logistic regression with different number of hashes for the (4-AV, 50\%) dataset. As observed, the more hash-functions used, the higher AUC achieved in the range of 70\% for 256 hashes and above, in line with~\cite{droidPermissionMiner}. In conclusion, the permissions set alone offers a moderate approach to detect Android malware. 

In the next sections we study the remaining 26 features (i.e. intrinsic, social and entity-related) along with 512 feature hashes and apply feature selection techniques to identify the most relevant ones.

\subsection{Feature selection}

Beginning at 538 features in the dataset, variable selection is performed to reduce model complexity. Generally, larger predictor collections do not necessarily imply better performance but larger complexity. In fact, the more predictors considered, the easier to bump into the well-known "Curse of dimensionality", which occurs when there is a large proportion of predictors with respect to data, penalising global performance.

\begin{figure*}[!htbp]
\centering
\subfigure[Features sorted by importance]{
\includegraphics[width=1\textwidth]{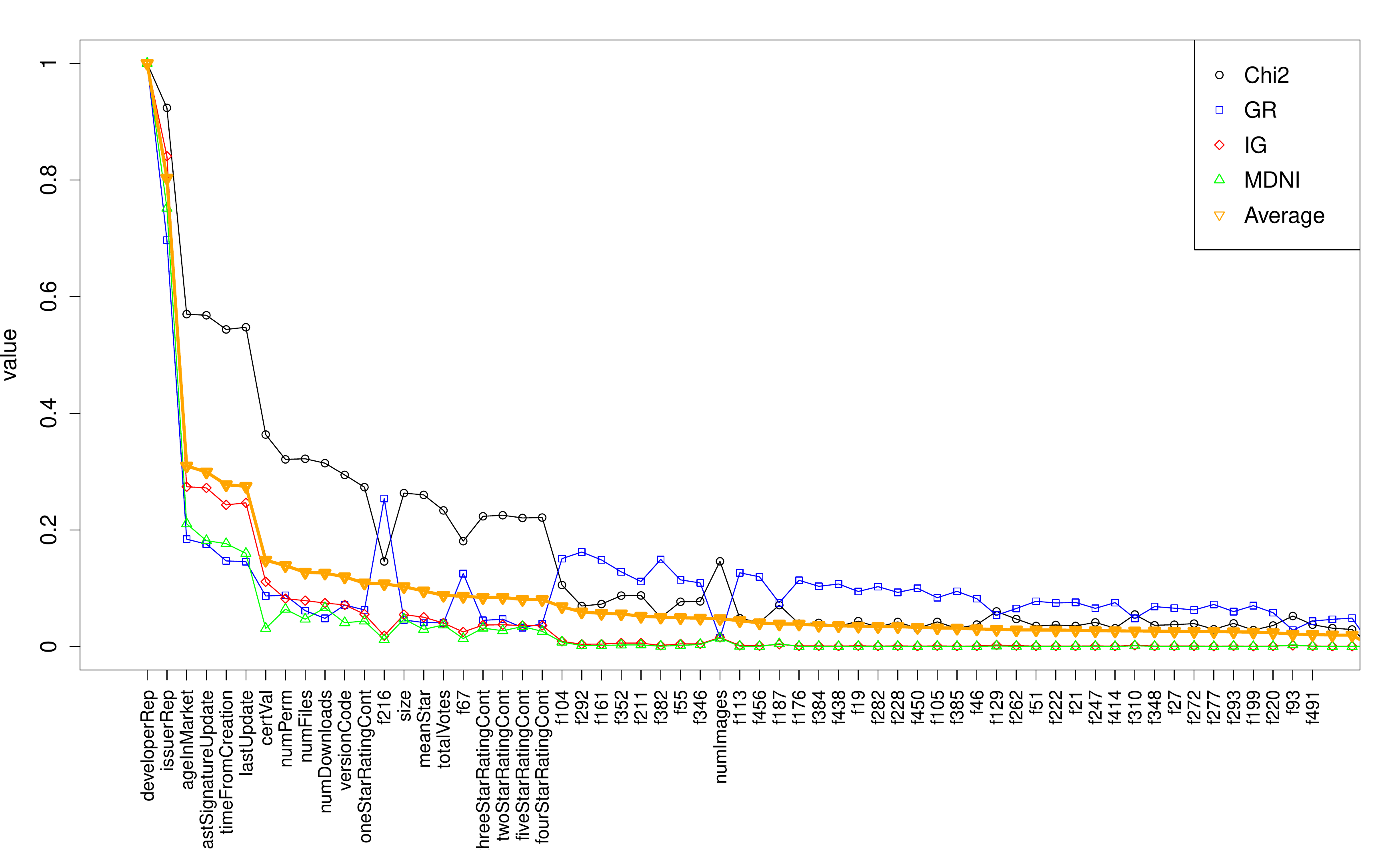}}
\subfigure[Performance of classifiers with different number of features]{
\includegraphics[width=1\textwidth]{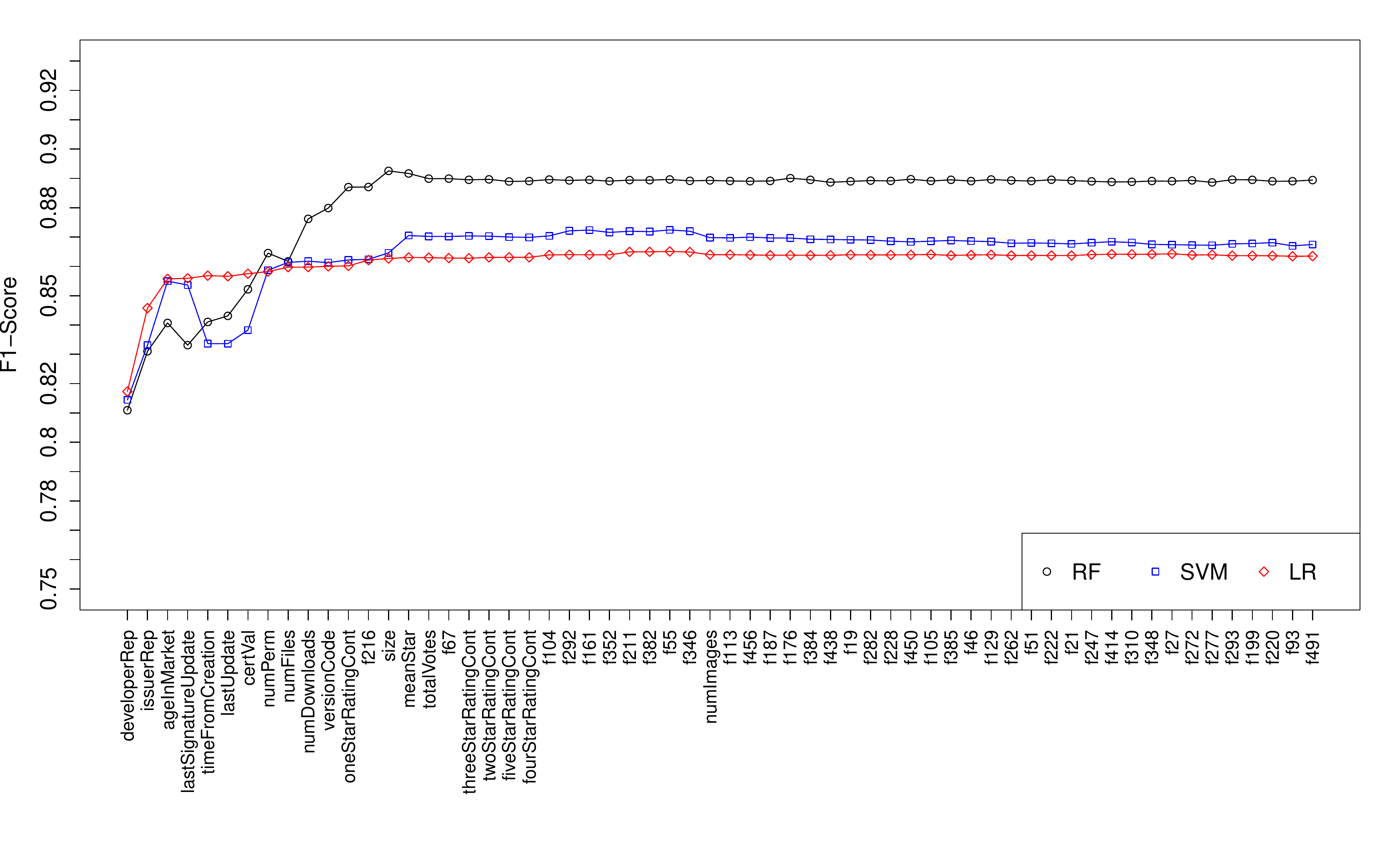}
}
\caption{Experiment results for feature selection}
\label{fig:variableSelection}
\end{figure*}

In the first experiment Fig.~\ref{fig:variableSelection} (top), we have used the four feature selection methods described in Section~\ref{sec:methodology} to evaluate the importance of each feature in the dataset. The results show such features sorted by each selection index and normalised with respect to the largest (names of features are self-explanatory). The dataset under study in this experiment was the (4-AV, 50\%).

As shown in Fig.~\ref{fig:variableSelection} (top), the top-7 most relevant features in the dataset are, in order of importance: developerRep, issuerRep, ageInMarket (number of days in market), lastSignatureUpdate, timeForCreation, lastUpdate and certVal. In contrast, the feature hashes on the permissions are not relevant when compared with the others. 

In order to establish the number of valid features for modeling, Fig.~\ref{fig:variableSelection} (bottom) shows the ten-fold cross-validated F1-score versus the number of predictors involved for each algorithm (RF, LR and SVM), where new predictors are added at each iteration in decreasing order of relevance. There, Random Forest provides the highest F1-score (around 0.89), while LR and SVM reach around 0.86 and 0.87 respectively. Moreover, the figure shows that highest performance on any algorithm may be achieved with only the top-15 features, which is set as predictor threshold.

In addition, it is worth remarking that developerRep alone achieves an F1-score above 0.8, showing that this metric alone is more powerful than any other, such as permissions.

\subsection{Malware detection model}

We perform a full benchmark test on the 9 composed datasets using only their top-15 features, namely: developerRep, issuerRep, ageInMarket (time in market), lastSignatureUpdate, timeForCreation, lastUpdate, certVal, numPerm, numFiles, numDownloads, versionCode, oneStarRatingCont, f216, size and meanStar. In this light, Table~\ref{tab:results} shows the training/test values of F1-score, precision and recall metrics for each dataset and the three models under study (LR, SVM, RF).

\begin{table}[!htbp]
\begin{tabular}{|  c c |  c c c |}
\hline
Malware & NumDetectors & F1-score & Precision  & Recall\\
\hline
\hline
\multicolumn{5}{|c|}{Logistic Regression (train/test)}   \\
\hline
2\% & 1  & 0.82/0.11&0.80/0.07&0.85/0.22\\
25\%  & 1 &0.89/0.62&0.93/0.61&0.85/0.63\\
50\%& 1 & 0.93/0.75&0.97/0.91&0.89/0.64\\
\hline
2\% & 2 &0.65/0.23&0.95/0.27&0.5/0.19 \\
25\%  & 2  &0.89/0.70&0.94/0.67&0.85/0.74 \\
50\%&  2  &0.94/0.83&0.98/0.9&0.90/0.76 \\
\hline
2\% & 4  &0.81/0.29&0.81/0.22&0.81/0.42\\
25\% &4  &0.91/0.76&0.95/0.72&0.87/0.79\\
50\%& 4  &0.95/0.86&0.99/0.86&0.92/0.86 \\
\hline
\hline
\multicolumn{5}{|c|}{Suport Vector Machine (train/test)}   \\
\hline
2\% & 1  &0.86/0.08&0.78/0.05&0.96/0.23 \\
25\%  & 1  &0.92/0.67&0.91/0.62&0.93/0.71 \\
50\%& 1 &0.95/0.81&0.96/0.88&0.94/0.76 \\
\hline
2\% & 2   &0.83/0.18&0.75/0.11&0.93/0.38  \\
25\%  & 2  &0.92/0.70&0.92/0.62&0.91/0.80\\
50\%& 2  &0.95/0.85&0.97/0.89&0.93/0.81\\
\hline
2\% & 4  &0.85/0.27&0.76/0.18&0.96/0.53 \\
25\%  &  4 &0.93/0.76&0.93/0.69&0.92/0.84\\
50\%& 4 &0.96/0.87&0.98/0.87&0.94/0.88\\
\hline
\hline
\multicolumn{5}{|c|}{Random Forest (train/test)}   \\
\hline
2\% & 1&0.99/0.12&0.99/0.08&0.99/0.32 \\
25\%  & 1 &0.99/0.73&0.99/0.70&0.99/0.76\\
50\%& 1 &0.99/0.83&0.99/0.87&0.99/0.8\\
\hline
2\% & 2  &0.99/0.22&0.99/0.15&0.99/0.45\\
25\%  & 2 &0.99/0.78&0.99/0.74&0.99/0.82 \\ 
50\%& 2  &0.99/0.87&0.99/0.89&0.99/0.85\\
\hline
2\% & 4  &0.99/0.32&0.99/0.22&0.99/0.58 \\
25\%  & 4  &0.99/0.82&0.99/0.77&0.99/0.86\\
50\%& 4 &0.99/0.89&0.99/0.88&0.99/0.90 \\
\hline
\end{tabular}
\caption{Full benchmark test with top-15 predictors.}
\label{tab:results}
\end{table}

\begin{table*}[!htb]
\centering
\begin{tabular}{|  c | c c c c c c c |}
\hline
\multicolumn{8}{|c|}{F1-score Random Forest (train/test)}   \\
\hline
 NDet  & 1-15 feats. & 3-17 feats.& 5-19 feats & 7-21 feats & 9-23 feats. & 11-25 feats & 13-27 feats \\
\hline
1 AV & 0.99/0.83 &0.99/0.86& 0.99/0.84 & 0.99/0.74& 0.96/0.72& 0.88/0.67 & 0.80/0.64\\
2 AV & 0.99/0.87 &0.99/0.87& 0.99/0.86 &0.99/0.79& 0.96/0.75& 0.88/0.71 &  0.75/0.66\\
4 AV & 0.99/0.89 &0.99/0.88 & 0.99/0.87 &0.99/0.80& 0.96/0.77& 0.89/0.73 & 0.77/0.69 \\ 
\hline
\end{tabular}
\caption{F1-score value of Random Forests with different feature sets.}
\label{tab:results2}
\end{table*}
The results show that algorithms achieve similar results, slightly better in the case of RF. Second, it might be observed that general performance improves as the percentage of malware samples increases, showing best results when malware accounts for 50\% of the applications. Actually, in the 2\%-malware case, the difference between train and test error suggests that the algorithms are overfitting the data. Finally, the algorithms perform best at identifying those malware applications tagged by several AV engines. Clearly, when the algorithms are trained with malware applications tagged by two engines or more, they reach up to 0.87 F1-score in the test set (bottom line in the table), thus providing a high-level prediction confidence.

\subsection{Robustness of the model}

The reader must note that malware developers, after reading this article, may decide to use different email accounts and certificates to evade this detection mechanism. However, the “malwarish” behaviour of applications is fingerprinted in several features redundantly. On the one hand, such redundancy implies that after 13-15 features no extra predictive power is gained by adding new features (as shown in Fig.~{\ref{fig:variableSelection}}); but on the other hand, it also provides robustness to the model since, if some features are decided not to be used (like developerRep and issuerRep) the others are still able to reach good performance.

To show this, Table~{\ref{tab:results2}} shows the F1-score results of re-running the RF algorithm to a different set of features. Essentially, the first column shows the same train/test F1-score values as in Table~{\ref{tab:results}} since both use the same top-15 features. The second column shows the F1-values when training and testing with features from 3 to 17 of Fig.~{\ref{fig:variableSelection}} (i.e. top-15 without developerRep and issuerRep). As shown the F1-score value is slightly worse than before. Similarly, when using features 5-19, a small decrease is observed again, but still good performance is achieved. F1-score quickly drops when using the features from position 7 in the ranking and on.

\section{Summary and Discussion}
\label{sec:conclusions}

In summary, this work has shown that Google Play meta-data provides valuable information to detect Android malware applications, reaching F1-score values near 0.9, for example when feeding meta-data to a Random Forest. In particular, it has been shown that using no more than 15 features, malware applications can be accurately identified.   

Furthermore, this work has also shown that inherent features, in particular application permissions, offer moderate prediction power (AUC-ROC about 0.7) compared to other metadata, such as the developer's reputation (percentage of malware applications uploaded by the same developer in the past) or certificate issuer reputation. This allows constructing efficient classification models for the early detection of malware applications uploaded at an Android market, as a prior step to more sophisticated techniques, namely code inspection or sandboxing.

The results of this works enable the use of simple static analysis at once for large amounts of Android applications. For apps uploaded to an application market, it might be determined whether it needs further inspection or it is suitable for direct upload. In addition, it is also possible to develop an in-device system which informs users about the appearance of each application and the risk of installing them in the device beforehand.

In a nutshell, the contributions of this work are the following:

\begin{itemize}
\item We evaluated the capabilities of permission-based detection approaches and their limitations by means of the hashing trick as feature reduction technique.
\item We showed that inherent application features, such as the developer's reputation (percentage of malware applications uploaded by the same developer in the past) or certificate issuer's reputation offer very good performance for detecting Android malware.
\item We proposed a model for Android Malware detection based on meta-data and machine learning techniques capable of detecting most Android threats, which can be leveraged both at market level and in-device application analysis.
\item We evaluated our proposed model over different benchmarking tests for performance and robustness of the algorithm
\end{itemize}

%techniques, namely code inspection or sandboxing.Finally, this methodology is shown to be robust since removing part of the features does not have a dramatic impact regarding accuracy and recall.

%Future work will attempt to improve malware detection performance by increasing the length and wide of data samples, including text-related features extracted from the user’s comments (text analytics). In addition, further research on malware engines will be undertaken as certain patterns have been observed among them.

\section*{Acknowledgements}

The authors would like to acknowledge the support of the Spanish project TEXEO (grant no. TEC2016-80339-R) and the EU-funded H2020 TYPES project (grant no. H2020-653449).

Additionally, Ignacio Mart\'{i}n would like to acknowledge the support of the Spanish Education Ministry for his FPU grant (grant no. FPU15/03518) which supports his position at UC3M.

\bibliographystyle{IEEEtran}
\bibliography{all}
\end{document}